\documentclass{osa-article}

\journal{optica}

\articletype{Research Article}

\usepackage{lineno}

\usepackage{siunitx}

\begin{document}

\title{Plug-\&-play generation of non-Gaussian states of light at a telecom wavelength}

\author{Mohamed F. Melalkia,\authormark{1,$\dagger$} Tecla Gabbrielli,\authormark{2,3,$\dagger$} Antoine Petitjean,\authormark{1} Léandre Brunel,\authormark{1} Alessandro Zavatta,\authormark{2,3} Sébastien Tanzilli,\authormark{1} Jean Etesse,\authormark{1} and Virginia D'Auria\authormark{1,4,*}}

\address{\authormark{1}Université Côte d'Azur, CNRS, Institut de Physique de Nice, Parc Valrose, 06108 Nice Cedex 2, France\\
\authormark{2}European Laboratory for Non-linear Spectroscopy (LENS), University of Florence, Via nello Carrara 1, 50019 Sesto Fiorentino (FI), Italy\\
\authormark{3}Istituto Nazionale di Ottica (CNR-INO), CNR, Largo Enrico Fermi 6, 50125 Firenze, Italy\\
\authormark{4}Institut Universitaire de France, France}

\email{\authormark{*}virginia.dauria@univ-cotedazur.fr} 



\begin{abstract*}
This work marks an important progress towards practical quantum optical technologies in the continuous variable regime, as it shows the feasibility of experiments where non-Gaussian state generation entirely relies on plug-\&-play components from guided-wave optics technologies. This strategy is demonstrated experimentally with the heralded preparation of low amplitude Schrödinger cat states based on single-photon subtraction from a squeezed vacuum. All stages of the experiment are based on off-the-shelf fiber components. This leads to a stable, compact, and easily re-configurable realization, fully compatible with existing fibre networks and, more in general, with future out-of-the-laboratory applications.
\end{abstract*}


\section{Introduction}

Optical quantum technologies in continuous variable (CV) regime find applications in quantum metrology \cite{schnabel2017squeezed, wang2020continuous}, quantum communication \cite{braunstein2005quantum, dias2020quantum, eberle2013gaussian}, and quantum computing \cite{asavanant2022optical, andersen2010continuous}. At the same time, many CV quantum information protocols requires the ability of generating non-Gaussian quantum states, such as Fock \cite{morin2012high} and Schrödinger cat-like states \cite{ourjoumtsev2007generation}, that are essential to fault-tolerant optical quantum information processing \cite{niset2009no} or long distance quantum communication \cite{takahashi2010entanglement, tipsmark2013displacement}. A common strategy to benefit from these resources is to implement conditional state preparation, often starting from a (Gaussian) squeezed state~\cite{lvovsky2020production}. This idea has been demonstrated in multiple CV experimental realizations~\cite{ ourjoumtsev2007generation,morin2012high, asavanant2017generation}, including the recent trend of employing a nonlinear waveguide as squeezing source~\cite{Takase:22}. Reported realizations, however, rely entirely, or largely, on bulk-optics setups that are generally hardly scalable and require careful spatial alignments~\cite{lvovsky2020production}. 

This work addresses the need for compact and stable realizations that can be easily assembled out-of-the-laboratory, as demanded to push current CV quantum optical experiments towards practical demonstrations. More specifically, it shows the feasibility of non-Gaussian state generation setups fully employing plug-\&-play components from guided-wave technologies. Strategically, we applied this concept to single-photon subtraction from a squeezed vacuum state. Such a scheme, commonly used for the preparation of small amplitude Schrödinger cat preparation~\cite{lvovsky2020production}, can be considered as an archetypal configuration, gathering Gaussian and non-Gaussian resources and lying at the heart of multiple complex architectures\cite{lvovsky2020production}. Squeezed states at a telecom wavelength are generated by single-pass spontaneous parametric down conversion (SPDC) in a commercial periodically poled lithium niobate ridge waveguide (PPLN/RW), whose input and output are fiber-pigtailed. Both photon subtraction operation and heralded state characterization via homodyne detection (HD) are then implemented using standard fiber components from classical telecom technology. The clear non-Gaussian behaviour of the measured heralded state proves the pertinence of our approach. This result opens the way to robust CV optical realizations compatible with existing telecommunication fiber networks and with considerably reduced resource overhead, spatial alignment, and mode matching issues.

\section{Experimental setup}

\begin{figure}[htbp]
\centering
\includegraphics[width=12cm]{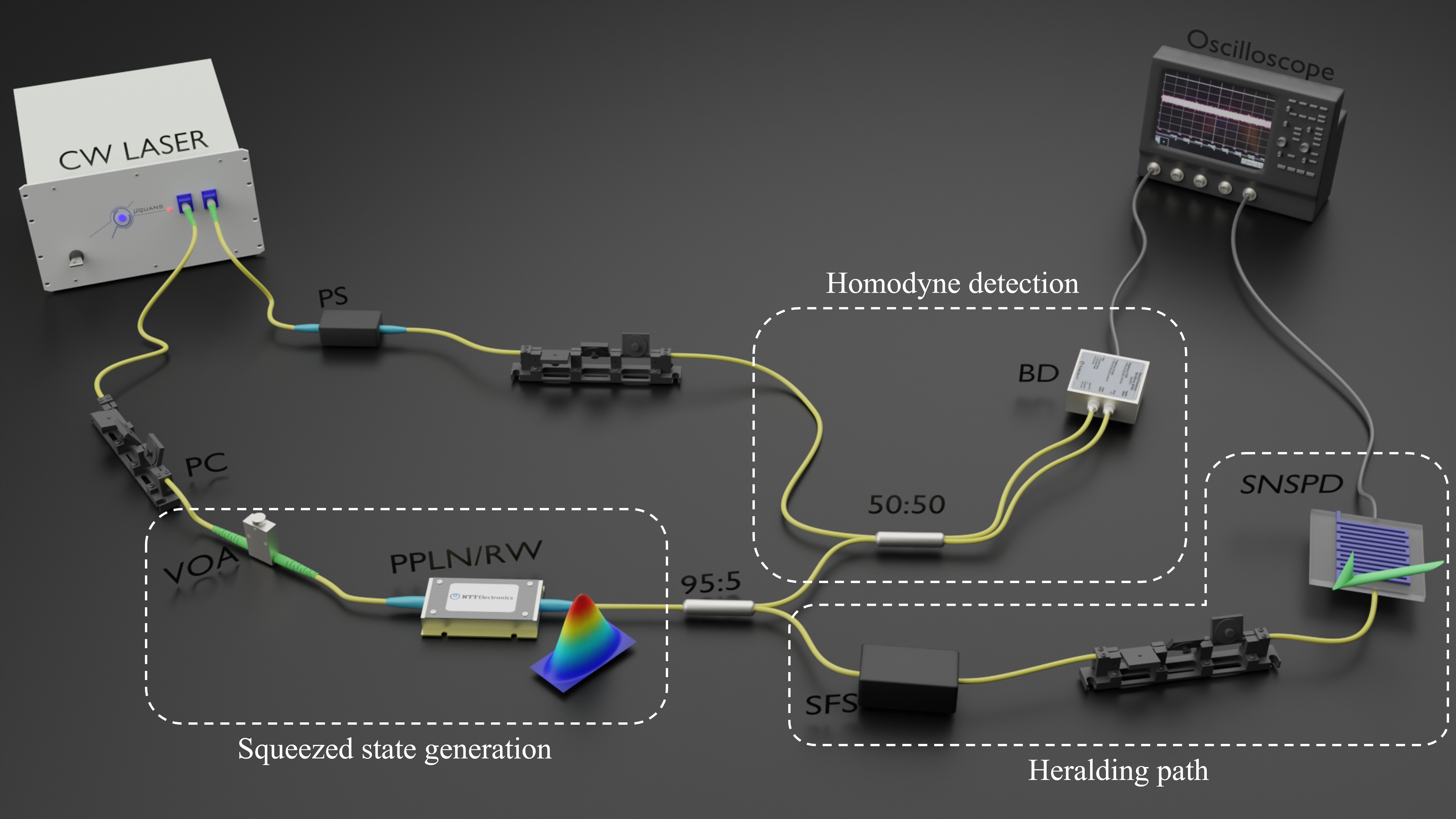}
\caption{Experimental setup. A fiber-coupled continuous wave (CW) laser delivers two outputs at $\lambda = \SI{1560.44}{nm}$ and $\lambda = \SI{780.22}{nm}$. The $\lambda = \SI{780.22}{nm}$ output passes through a polarization controller (PC) and a variable fiber optical attenuator (VOA) to be used as pump source for the PPLN/RW generating the single-mode squeezed vacuum. Downstream of the PPLN/RW, squeezed light is sent to a 95:5 fiber beam-splitter. The $5\%$ output undergoes a spectral filtering stage (SFS) and is sent to a superconducting nanowire single-photon detector (SNSPD), whose signal heralds the non-Gaussian state generation. The $95\%$ output is sent to a homodyne detection (HD) made of a 50:50 fiber beam-splitter and a commercial balanced detector (BD). The HD local oscillator is given by the laser output at $\lambda = \SI{1560.44}{nm}$, passing through a fiber phase shifter (PS) and a polarization controller (PC).}\label{fig:exp_setup}
\end{figure}

Fig. \ref{fig:exp_setup} shows the detailed experimental setup. The master source is a continuous wave (CW) fiber-coupled laser (Muquans) delivering outputs at $\lambda = \SI{1560.44}{nm}$ and its second harmonic (SH) at half of the wavelength ($\lambda = \SI{780.22}{nm}$). The SH output is sent to fiber polarization controller (PC) and an optical variable attenuator (VOA) that allow adjusting its polarization and intensity. This set of plug-\&-play telecom components is then directly connected to the PPLN/RW (NTT Electronics corp). Single-mode squeezed vacuum state at $\lambda = \SI{1560.44}{nm}$ is generated via type-0 frequency degenerate SPDC and collected by the optical fiber at the PPLN/RW output with an efficiency $\eta_{c}$. Propagation losses inside the nonlinear waveguide lead to a reduced PPLN/RW transmission efficiency $\eta_{pl}$. As largely documented in the literature~\cite{lvovsky2020production}, heralded single-photon subtraction can be implemented by sending squeezed light to a low reflectivity beam-splitter followed by a single-photon detection stage. In our experiment, a 95:5 fiber beam-splitter (f-BS) splits squeezed light into two paths: the $5\%$ output is sent to the heralding path, whereas the $95\%$ output, carrying the heralded non-Gaussian state, goes to the HD for the measurement of its CV properties. 
Following an approach that we already validated for squeezing experiments~\cite{kaiser2016fully}, the HD is realized by means of a 50:50 f-BS whose outputs are connected to a commercial balanced detection (Insight) showing a bandwidth of $\SI{300}{MHz}$. The HD local oscillator (LO) is provided by the fundamental laser output ($\lambda = \SI{1560.44}{nm}$), whose phase, $\theta$, is controlled via a fiber phase shifter. The overall detection efficiency is $\eta_{HD} = \eta_{t}\cdot\eta_{pd}\cdot\eta_{el} = 0.72$, where $\eta_{t}= 0.94$ includes measured losses in the f-BSs, $\eta_{pd}=0.80$ is the photodiode efficiency as given by the manufacturer, and $\eta_{el}=0.96$ accounts to the residual HD electronic noise. The HD output is monitored using an oscilloscope (Teledyne LeCroy) for data post-processing. The CV data acquisition on the oscilloscope is triggered by photon counting events from the heralding path.\\   
Light on the heralding path undergoes a cascade of optical spectral filters allowing to comply with the strong mismatch between the HD electronical bandwidth ($\SI{300}{MHz}$) and the broad squeezing optical emission bandwidth (a few THz). Indeed, photon-counting events of wavelengths far away from the CW LO could trigger the CV detection of signals whose non-Gaussian features are out-of-the-reach of HD electronics. This would weave data corresponding to the desired non-Gaussian states with those relative to squeezed vacuum, eventually resulting in the measurement of a mixed state. Such an effect further combines with the fact that single-pass squeezing can be not pure over its whole bandwidth~\cite{yoshikawa2017purification}. Spectral filtering in the heralding path allows circumventing these imperfections~\cite{yoshikawa2017purification, melalkia2022theoretical}. In our realization, the $5\%$ of tapped light from the squeezed state is filtered using a commercial $\SI{500}{MHz}$ FWHM fiber Bragg grating filter (FBG from AOS GmbH) and an in-house fiber-coupled Fabry-Pérot cavity with a target bandwidth of $\SI{10}{MHz}$. Note that, as an alternative, plug-\&-play optical filters with bandwidths of a few tens of MHz are commercially available. Filtered light is then sent to a polarization controller and to a superconducting nanowire single-photon detector (SNSPD from IDQuantique). The heralding rate is of $\SI{3}{kHz}$ and the dark count rate is about $\SI{80}{Hz}$, giving a signal to noise ratio, SNR, of $\SI{15.7}{dB}$. 

\section{Data analysis}
\begin{figure}[htbp]
\centering
\includegraphics[width=6cm]{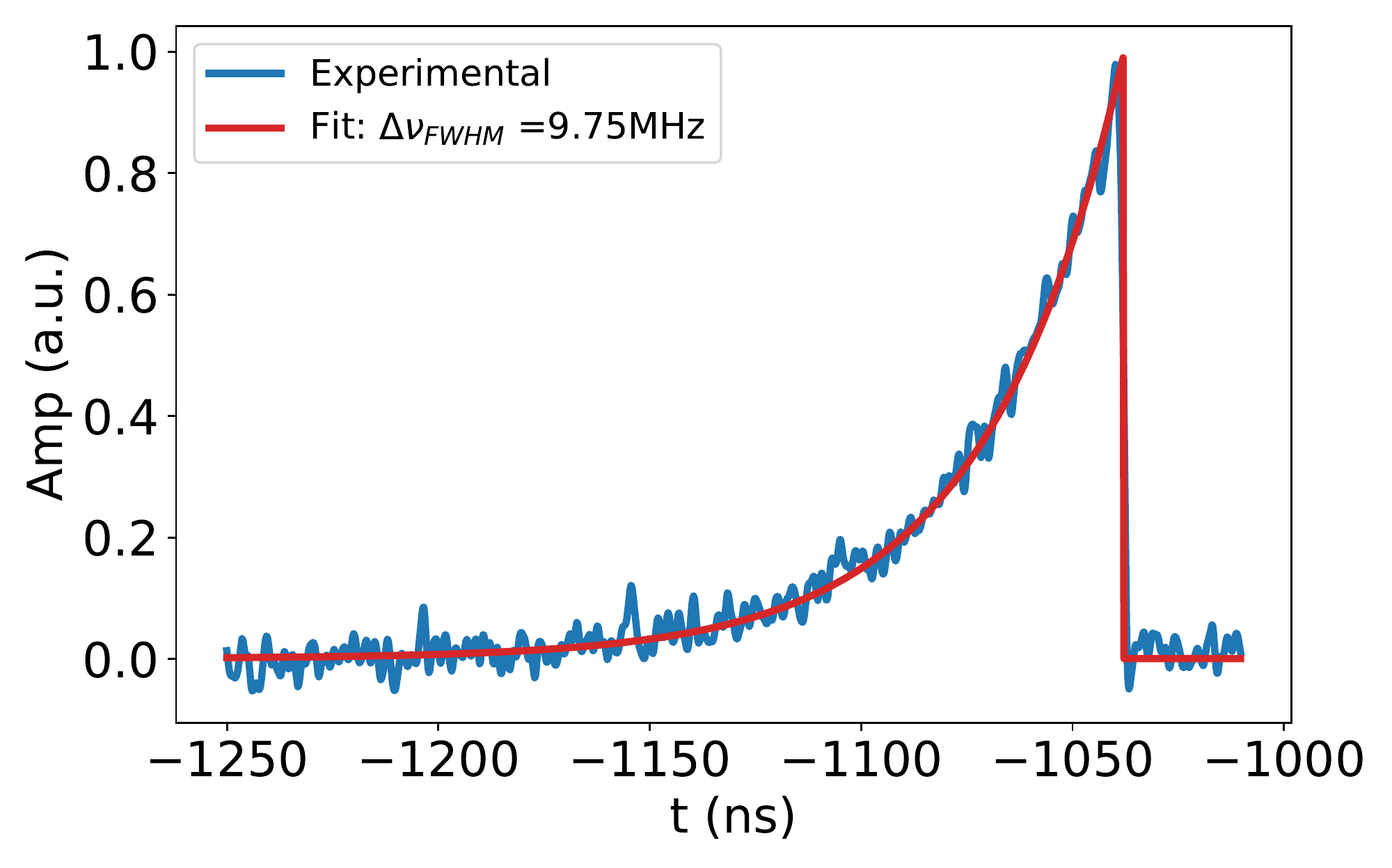} 
\caption{The temporal mode function corresponding to the highest eigenvalue (blue curve) and the fit with the theoretical model (red curve) as functions of time. The cascade of the Fabry-Pérot cavity and the fiber Bragg grating filter in the heralding path gives a filtering bandwidth of $\SI{9.75}{MHz}$ at FWHM.} \label{fig:eig_mode}
\end{figure}

The properties of the heralded non-Gaussian state can be reconstructed by performing a quantum state tomography, starting from the state quadratures measured by the HD~\cite{leonhardt1997measuring}. A series of $43000$ homodyne traces, $x_i(t)$, of a duration of $\SI{5}{\mu s}$ each at a sampling rate of $\SI{5}{Gs/s}$ are recorded using the oscilloscope. Note that the LO phase is spanned with a speed of approximately $9\pi / s$ and, therefore, it can be considered as a constant over the time needed to acquire a homodyne trace. For each acquisition $x_i(t)$, the actual $\theta_i$ value is inferred from the phase dependency of the measured squeezed vacuum variance. \\
As discussed in~\cite{yoshikawa2017purification}, in the CW regime, the quadrature $X_i$ of the heralded non-Gaussian state can be obtained from raw homodyne data as~\cite{lvovsky2015squeezed}:

\begin{equation}\label{datiraw}
X_i = \int f(t) x_i(t) dt,
\end{equation}

\noindent where $f(t)$ is the temporal mode of the heralded non-Gaussian state, as determined by the filtering stages on the heralding path. When the squeezing bandwidth is much broader than the bandwidth of the filtering cavity, $\gamma / \pi$, the theoretical $f(t)$ is expected to be $\sim e^{\gamma t} u(-t)$, $u(t)$ being the Heaviside step function. Experimentally, $f(t)$ is obtained as in~\cite{morin2013experimentally} by computing the autocorrelation function of the heralded homodyne traces, $\langle x(t)x(t')\rangle$, and by numerically performing its single value decomposition: the eigenfunction corresponding to the highest eigenvalue gives the optimal $f(t)$. Fig. \ref{fig:eig_mode} shows both the reconstructed temporal mode $f(t)$ (blue) and theoretical one (red); a perfect agreement between the two curves can be observed and gives a filtering bandwidth of $\gamma / \pi = \SI{9.75}{MHz}$ at FWHM, in agreement with the filtering stage characteristics.

 By exploiting Eq.~\eqref{datiraw}, the obtained $f(t)$ is used to extract the quadratures of the squeezed state and of the heralded non-Gaussian state out of the raw homodyne data. More in details, with reference to Fig. \ref{fig:eig_mode}, $x_i(t)$ data are post-processed within the time window $\approx [\SI{-1.25}{\mu s}, \SI{-1}{\mu s}]$ to extract the heralded non-Gaussian state quadratures. Data outside this time window are used to extract the squeezed state quadrature. 

\section{Experimental results and discussion}

The tomographic analysis allows measuring, for the non-heralded state, a squeezing of $-1.80 \pm 0.05$ dB and an antisqueezing of $3.36 \pm 0.05$ dB. This corresponds to an initial squeezing of $-5.39 \pm 0.05$ dB when corrected for the total losses experienced by squeezed light, $\eta_{tot} = \eta_{wg}\cdot\eta_s\cdot\eta_{HD} = 0.48$. In the expression of $\eta_{tot}$, $\eta_s = 0.96$ is the transmission of the subtracting 95:5 f-BS and $\eta_{wg}$ accounts for both the propagation losses and the guide-to-fibre coupling occurring at the SPDC stage ($\eta_{wg} = \eta_{pl}\cdot\eta_{c}$). From the measured $\eta_s$ and $\eta_{HD}$, we deduce $\eta_{wg} = 0.69$. \\

\begin{figure}[htbp]
\centering
\begin{tabular}{cc}
(a) & (b) \\ 
\includegraphics[width=6cm]{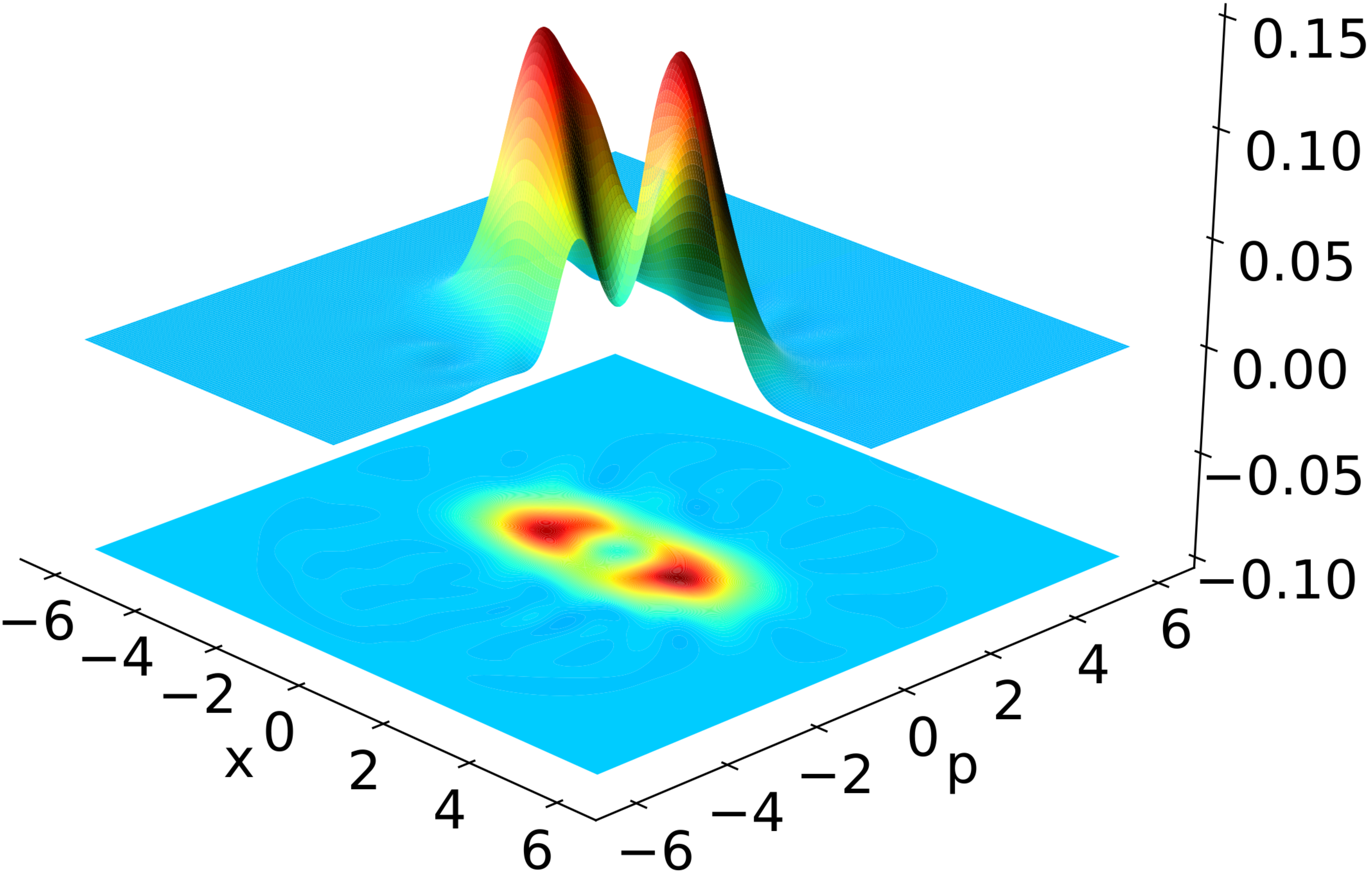} & \includegraphics[width=6cm]{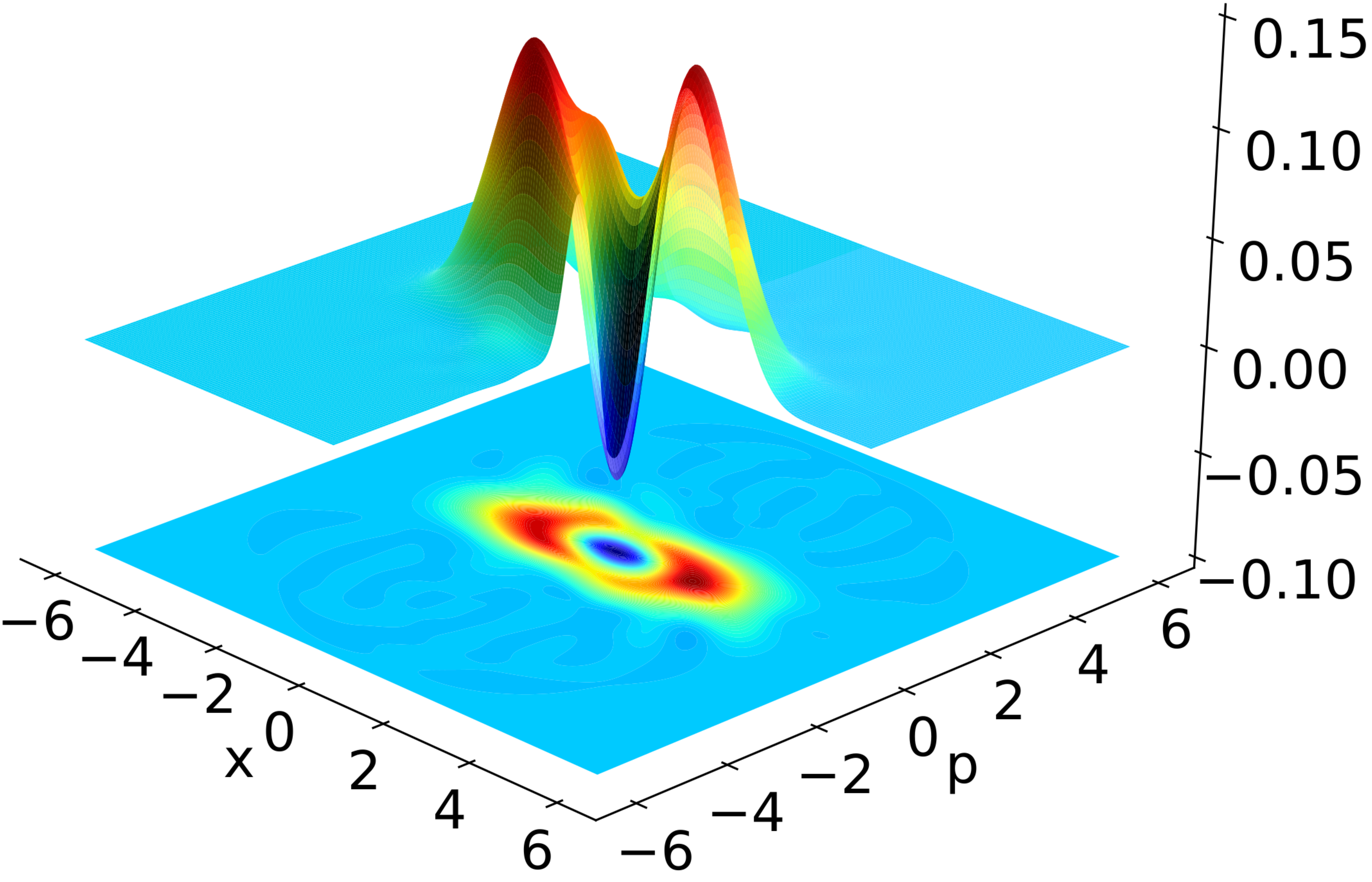} \\ 
\end{tabular} 
\caption{(a) The Wigner function $W_{NG}(x, p)$ of the heralded non-Gaussian state and (b) the Wigner function $W_{Corr}(x, p)$ of the state corrected by the homodyne detection efficiency $\eta_{HD} = 0.72$.} \label{fig:Wigner}
\end{figure}

The tomography of the heralded non-Gaussian state is performed with the maximum-likelihood algorithm with $200$ iterations \cite{lvovsky2009continuous}. Fig. \ref{fig:Wigner}-(a) shows the obtained Wigner function of the generated non-Gaussian state. As can be seen, the generated state has a highly non-Gaussian Wigner function with a minimum at phase origin $W_{NG}(0, 0) = 0.016 \pm 0.004$. Its profile perfectly matches the one predicted by theory ($W_{Th}(0, 0) = 0.015$) following an approach similar to the one in \cite{ourjoumtsev2007etude, morin2013non} and taking into account all the experimental imperfections. This result represents the very first demonstration of non-Gaussian states in CV regime obtained using a fully guided-wave approach and it attests its impact towards compact, simple, and plug-\&-play setups involving both Gaussian and non-Gaussian operations. 

In this regard, we observe that without modifying $\eta_{wg}$, which is in agreement with realistic performances in such fibre-coupled ridge waveguides, our results could be straightforwardly enhanced by simply replacing our  balanced photodetection with one with higher efficiency. By correcting the data by the HD efficiency ($\eta_{HD}= 0.72$)~\cite{leonhardt1997measuring}, the non-Gaussian state obtained just after the 95:5 f-BS exhibits a negative Wigner function as expected for Schrödinger cat states, the corresponding value at the phase origin being $W_{Corr}(0, 0) = -0.065 \pm 0.004$ (see Fig. \ref{fig:Wigner}-(b)). We point out that residual losses in the fiber components and low HD efficiency are not intrinsic to the followed guided-wave approach but only to our specific realization. In particular, the lack of efficient photodiodes, that represent the most critical loss factor in the HD, affects in the same way both guided-wave and bulk realizations. 

As concluding remark, we stress that a further improvement of the Wigner negativity could be obtained by improving $\eta_{wg}$. The combined effect of propagation and coupling losses at the nonlinear waveguide leads to a degradation of the purity of the squeezed state on which the subtraction protocol is applied. This can be circumvented by adopting a full integrated setup including squeezed state generation, photon subtraction and homodyne detection, in a configuration similar to the one shown for Gaussian state generation, manipulation, and detection on lithium niobate circuits~\cite{mondain2019chip, lenzini2018integrated, zhao2020near}. Such a strategy would push the propagation losses down to $\SI{0.03}{dB/cm}$ and the coupling losses to 0, thus giving for a typical chip length of $\SI{5}{cm}$, a total $\eta_{wg} = 0.97$. In this condition, the Wigner function negativity would go to its maximal value, $W_{Corr}(0, 0) = -0.22$.

\section{Conclusion}

We have demonstrated that a plug-\&-play, fully guided-wave, approach to CV experiments is definitely reachable with off-the-shelves standard components and devices. This concept has been pushed, for the first time, to the generation of non-Gaussian states of light, showing the possibility of implementing complex setups free from any bulk optics stage. The obtained realization enables  reconfigurable and scalable systems compatible with integrated photonics~\cite{mondain2019chip, lenzini2018integrated, zhao2020near}, adapted to ambitious CV quantum communication protocols. These ingredients are essential to scale-up realizations of increasingly complex optical quantum systems capable of enabling disruptive but realistic quantum information protocols. 

\begin{backmatter}
\bmsection{Funding}
This work has been conducted within the framework of the project OPTIMAL granted by the European Union by means of the "Fond Européen de développement regional" (FEDER). The authors also acknowledge financial support from the Agence Nationale de la Recherche (ANR) through the project HyLight (ANR-17-CE30-0006-01) and the French government through its program "Investments for the Future" under the Université Côte d'Azur UCA-JEDI project managed by the ANR (grant agreement ANR-15-IDEX-01).

\bmsection{Acknowledgments}
Virginia D'Auria acknowledges financial support from the Institut Universitaire de France (IUF).

\bmsection{Disclosures}
The authors declare no conflicts of interest.

\bmsection{Data availability} Data underlying the results presented in this paper are not publicly available at this time but may be obtained from the authors upon reasonable request.\\
\end{backmatter}

\noindent \authormark{$\dagger$}These authors contributed equally to this work.


\bibliography{Bibliography}

\end{document}